\documentstyle[prl,aps,epsf,preprint]{revtex}

  \let\mysection=\section
  \renewcommand{\section}{\setcounter{equation}{0}\mysection}

\def\gl{\,\hbox{\rlap{\raise.5ex\hbox{$>$}}\lower.5ex\hbox{$<$}}\,}

\def\upbracketfill{
       $\leaders\hrule\hfill$}
\def\underbracket#1{\mathop{\vtop{\ialign{##\crcr
   $\hfil\displaystyle{#1}\hfil$\crcr\noalign{\kern3pt\nointerlineskip}
   \upbracketfill\crcr\noalign{\kern3pt}}}}\limits}
\def\upslurfill{
       $\bracelu\leaders\vrule\hfill\braceru$}
\def\underslur#1{\mathop{\vtop{\ialign{##\crcr
   $\hfil\displaystyle{#1}\hfil$\crcr\noalign{\kern3pt\nointerlineskip}
   \upslurfill\crcr\noalign{\kern3pt}}}}\limits}

\def\bop#1{\setbox0=\hbox{$#1M$}\mkern1.5mu
	\vbox{\hrule height0pt depth.04\ht0
	\hbox{\vrule width.04\ht0 height.9\ht0 \kern.9\ht0
	\vrule width.04\ht0}\hrule height.04\ht0}\mkern1.5mu}
\def\bo{{\mathpalette\bop{}}}                        

\def\llsim{\:\lower1.8ex\hbox{$\buildrel<\over{\widetilde{\phantom{
<}}}$}\:}

\def\boxes#1{
        \newcount\num
        \num=1
        \newdimen\downsy
        \downsy=-1.5ex
        \mskip-2.8mu
        \bo
        \loop
        \ifnum\num<#1
        \llap{\raise\num\downsy\hbox{$\bo$}}
        \advance\num by1
        \repeat}
\def\boxup#1#2{\newcount\numup
        \numup=#1
        \advance\numup by-1
        \newdimen\upsy
        \upsy=.75ex
        \mskip2.8mu
        \raise\numup\upsy\hbox{$#2$}}

\begin{document}
\font\mybbb=msbm10 at 8pt
\font\mybb=msbm10 at 12pt
\def\bbb#1{\hbox{\mybbb#1}}
\def\bb#1{\hbox{\mybb#1}}
\def\id{\protect{{1 \kern-.28em {\rm l}}}}
\def\I {\bb{1}}
\def\Z {\bb{Z}}
\def\pRe{\bbb{R}}
\def\Re {\bb{R}}
\def\C {\bb{C}}
\def\pC{\bbb{C}}
\def\H {\bb{H}}
\def\ni{\noindent}
\newcommand\noi{\noindent}
\def\nn{\nonumber}
\newcommand\seq{\;\;=\;\;}
\def\de{\nabla}
\def\imp{~~\Rightarrow~~}

\def\eq{\begin{equation}}
\def\eqe{\end{equation}}
\def\eqa{\begin{eqnarray}}
\def\eqae{\end{eqnarray}}
\def\bea{\begin{eqnarray}}
\def\ena{\end{eqnarray}}

\def\st{\star}
\def\dZ2p{\frac{dZ_1}{2\p i}}

\def\half{\frac{1}{2}}
\def\({\left(} \def\){\right)} \def\<{\langle } \def\>{\rangle }
\def\[{\left[} \def\]{\right]} \def\lb{\left\{} \def\rb{\right\}}

\newcommand{\be}{\begin{equation}}
\newcommand{\ee}{\end{equation}}
\newcommand{\ba}{\begin{eqnarray}}
\newcommand{\ea}{\end{eqnarray}}
\newcommand{\al}{\mbox{$\alpha$}}
\newcommand{\als}{\mbox{$\alpha_{s}$}}
\newcommand{\s}{\mbox{$\sigma$}}
\newcommand{\lm}{\mbox{$\mbox{ln}(1/\alpha)$}}
\newcommand{\bi}[1]{\bibitem{#1}}
\newcommand{\fr}[2]{\frac{#1}{#2}}
\newcommand{\sv}{\mbox{$\vec{\sigma}$}}
\newcommand{\gm}{\mbox{$\gamma_{\mu}$}}
\newcommand{\gn}{\mbox{$\gamma_{\nu}$}}
\newcommand{\Le}{\mbox{$\fr{1+\gamma_5}{2}$}}
\newcommand{\R}{\mbox{$\fr{1-\gamma_5}{2}$}}
\newcommand{\GD}{\mbox{$\tilde{G}$}}
\newcommand{\gf}{\mbox{$\gamma_{5}$}}
\newcommand{\om}{\mbox{$\omega$}}
\newcommand{\Ima}{\mbox{Im}}
\newcommand{\Rea}{\mbox{Re}}
\newcommand{\Tr}{\mbox{Tr}}

\newcommand{\bbeta}[2] {\mbox{$\bar{\beta}_{#1}^{\hspace*{.5em}#2}$}}
\newcommand{\homega}[2]{\mbox{$\hat{\omega}_{#1}^{\hspace*{.5em}#2}$}}

\def\a{\alpha}
\def\b{\beta}
\def\c{\chi}
\def\d{\delta}
\def\e{\epsilon}           
\def\f{\phi}               
\def\g{\gamma}
\def\h{\eta}
\def\i{\iota}
\def\j{\psi}
\def\k{\kappa}                    
\def\l{\lambda}
\def\m{\mu}
\def\n{\nu}
\def\o{\omega}
\def\p{\pi}                
\def\q{\theta}                    
\def\r{\rho}                      
\def\s{\sigma}                    
\def\t{\tau}
\def\u{\upsilon}
\def\x{\xi}
\def\z{\zeta}
\def\D{\Delta}
\def\F{\Phi}
\def\G{\Gamma}
\def\J{\Psi}
\def\L{\Lambda}
\def\O{\Omega}
\def\P{\Pi}
\def\Q{\Theta}
\def\S{\Sigma}
\def\U{\Upsilon}
\def\X{\Xi}
\def\del{\partial}
\def\pa{\partial}


\def\ca{{\cal A}}
\def\cb{{\cal B}}
\def\cc{{\cal C}}
\def\cd{{\cal D}}
\def\ce{{\cal E}}
\def\cf{{\cal F}}
\def\cg{{\cal G}}
\def\ch{{\cal H}}
\def\ci{{\cal I}}
\def\cj{{\cal J}}
\def\ck{{\cal K}}
\def\cl{{\cal L}}
\def\cm{{\cal M}}
\def\cn{{\cal N}}
\def\co{{\cal O}}
\def\cp{{\cal P}}
\def\cq{{\cal Q}}
\def\car{{\cal R}}
\def\cs{{\cal S}}
\def\ct{{\cal T}}
\def\cu{{\cal U}}
\def\cv{{\cal V}}
\def\cw{{\cal W}}
\def\cx{{\cal X}}
\def\cy{{\cal Y}}
\def\cz{{\cal Z}}

\def\vecnab{\vec{\nabla}}
\def\vx{\vec{x}}
\def\vy{\vec{y}}
\def\arrowk{\stackrel{\rightarrow}{k}}
\def\kbar{k\!\!\!^{-}}
\def\karrow{k\!\!\!{\rightarrow}}
\def\arrowl{\stackrel{\rightarrow}{\ell}}
\def\var{\varphi}

\def\fig #1 by #2 (#3){\vbox to #2{
	\hrule width #1 height 0pt depth 0pt\vfill\special{picture #3}}}

%

\def\hook#1{{\vrule height#1pt width0.4pt depth0pt}}
\def\leftrighthookfill#1{$\mathsurround=0pt \mathord\hook#1
        \hrulefill\mathord\hook#1$}
\def\underhook#1{\vtop{\ialign{##\crcr                 
        $\hfil\displaystyle{#1}\hfil$\crcr
        \noalign{\kern-1pt\nointerlineskip\vskip2pt}
        \leftrighthookfill5\crcr}}}
\def\smallunderhook#1{\vtop{\ialign{##\crcr      
        $\hfil\scriptstyle{#1}\hfil$\crcr
        \noalign{\kern-1pt\nointerlineskip\vskip2pt}
        \leftrighthookfill3\crcr}}}
\def\claw#1{{\vrule height0pt width0.4pt depth#1 pt}}
\def\leftrightclawfill#1{$\mathsurround=0pt \mathord\claw#1
        \hrulefill\mathord\claw#1$}
\def\overhook#1{\vbox{\ialign{##\crcr\noalign{\kern1pt}
       \leftrightclawfill5\crcr\noalign{\kern1pt\nointerlineskip}
       $\hfil\displaystyle{#1}\hfil$\crcr}}}

\def\under#1#2{\mathop{\null#2}\limits_{#1}}

\def\boxit#1{\leavevmode\thinspace\hbox{\vrule\vtop{\vbox{
	\hrule\kern1pt\hbox{\vphantom{\bf/}\thinspace{\bf#1}\thinspace}}
	\kern1pt\hrule}\vrule}\thinspace}

\def\ttZ{\tilde{\tilde{Z}}}
\def\ttz{\tilde{\tilde{z}}}
\def\tz{\tilde{z}}
\def\tZ{\tilde{Z}}
\def\tq{\tilde{\q}}
\def\del{\partial}
\def\st{\star}
\def\stam{\stackrel{\rightarrow}{\m}}
\def\stab{\stackrel{\rightarrow}{\b}}
\def\half{\frac{1}{2}}
\def\m{\mu}
\def\n{\nu}
\def\g{\gamma}
\def\d{\delta}
\def\in{\infty}
\def\nn{\nonumber\\}
\baselineskip=24pt
\pagenumbering{arabic}

\font \biggbold=cmbx10 scaled\magstep3
\font \bigbold=cmbx10 at 12.5pt
\font \bigreq=cmr10 at 12 pt
\baselineskip=24pt
\hfill{}
\vskip .50cm
{\biggbold \centerline{On Some Positivity Properties of the }}
{\biggbold \centerline{Interquark Potential in QCD}}
\vskip 1.0cm
{\bigbold \centerline{Shmuel Nussinov\footnote{Electronic 
mail: nussinov@ccsg.tau.ac.il}}}
\vskip 1.0cm
{\baselineskip=16pt
\centerline{\it School of Physics and Astronomy }
\centerline{\it Raymond and Beverly Sackler Faculty of Exact Sciences  }
\centerline{\it Tel Aviv University}
\centerline{\it 69978 Tel Aviv, Israel}
\centerline{\it and}
\centerline{\it Department of Physics and Astronomy}
\centerline{\it University of South Carolina}
\centerline{\it Columbia, South Carolina  29208, USA}}
\vskip 2.0cm
{\bigbold \centerline{Abstract}}
\vskip 2.0cm
\baselineskip=16pt

We prove that the Fourier transform of the exponential $e^{-\b V(R)}$ 
of the {\bf static} interquark potential in
QCD is positive.  It has been shown by Eliott Lieb some time ago that 
this property allows in the same limit of static
spin independent potential proving certain mass relation between 
baryons with different quark flavors.

\newpage

Convex potentials and potentials with positive laplacians have bound 
states with radial/orbital excitations
ordered in certain ways \cite{ref:feldman} \cite{ref:baum}.  The 
convexity of $V(r)$ has been proven some time ago
\cite{ref:bachas}.

Here we would like to argue that the Fourier transform of $e^{-\beta 
V (r)}$ is positive.  E. Lieb \cite{ref:lieb} has
shown that this property implies, in particular the inequality (first 
conjectured in \cite{nu1}):
\eqa
M (mmm) + M (m \mu \mu) = 2 M (mm \mu)
\eqae
for baryons made of quarks with the above masses and with spin and 
flavor (mass) independent pairwise
interaction $V (r)$.

Following Wilson we define for any closed path $c$ in an Euclidean 
four-dimensional Lattice:
\eqa
W (c) = < tr Uc>
\eqae
with
\eqa
U_c = \prod_{l = {\rm links} \e c} U_l
\eqae
the path ordered product of the $U$ matrices associated with the 
links of $c$.  The Euclidean path integral average
is indicated by $<>$.  It involves integrating over all $d U_l$ with 
the positive weight:
\eqa
e^{- \sum\limits_{\rm plaquettes}  tr (U (p) + U^+ (p) -2)}
\prod_i \det S^{(i)}_F (U)
\eqae
with $(S^i_F (U)$  the fermion propagator in the $U$ background and 
can be shown to be
positive $\cite{ref:weingarten}, \cite{ref:vafanp}, \cite{ref:nnm}$. 
Consider the special case where $c=c (R,T)$ is a
rectangle of ``time-direction" height
$T$ and width $R$.  The static potential between two heavy quarks 
separated by a distance $R$ is given by
\eqa
V(R) = \lim\limits_{T - \infty} - {1 \over T} \ln W(R,T) \; .
\eqae
The argument for (4) is well-known.

Let us start at $t=0$ with a $Q \bar{Q}$ state with the quark $Q$ and 
anti-quark $\bar{Q}$ at $\pm {\vec{R} \over
2}$.  To maintain gauge invariance we should multiply the state by a 
Wilson line factor, $\Pi U_i$, containing the
product of the $U$'s on $c  ( - {\vec{R} \over 2} , {\vec{R} \over 
2})$, namely on the spatial links along some
  path connecting ${\vec{R} \over 2}$ and $-{\vec{R} \over 2}$. The 
latter is usually chosen to be the straight line of
length $R$ connecting $+ {\vec{R} \over 2}$ and
$- {\vec{R} \over 2}$.  We then let the system propagate for time 
$T$.  The (very) massive quarks do then stay at
the initial location and hence propagte over the straight ``pure 
time" lines $L_Q, L_{\bar{Q}}$ connecting $(+
{\vec{R} \over 2} , 0)$ with $(- {\vec{R} \over 2} , T)$ and $(+ 
{\vec{R} \over 2} , 0)$ with $(+ {\vec{R} \over 2} , T)$.
In the process
$Q$ picks up the Wilson factor $U_{L_1}$ associated with the upgoing 
$L_1$ and $\bar{Q}$ picks up $U^+_{L_2} =
U_{-L_2}$, the conjugate or line-reversed Wilson factor for $L_2$. 
The complete evolution of the system, given
by
\eqa
< \psi_{f \; Q \bar{Q}} \mid e^{i H T} \mid \psi_{i \; Q \bar{Q}} >
\eqae
will then have the four elements of $W(R,T)$ pieced together into one 
closed loop.  (Recall that $< \psi_{f \; Q
\bar{Q}} \mid$ includes the opposite Wilson line (from $- {\vec{R} 
\over 2} , \; {\rm to} \; {\vec{R} \over 2}$).)  This
becomes the matrix element of $e^{-HT}$ after Wick rotation to the 
Euclidean regime.  For static
quarks this is simply: $e^{-2MT} e^{-V(R)T}$.  It is calculable via 
the Euclidean langragian path integral leading to
\eqa
e^{(-V(R))T} \approx < tr U(R)T> = < W(R,T)> \; .
\eqae
where the factor $e^{-2MT}$ associated with the free propagation of 
the two heavy quarks is omitted, and
equation (5) readily follows.

We can separate $U(c) = U_L U_R^+$ into the product of all $U$'s 
lying on links to the
left (right) of, say the $x=0$ reflection plane and correspondingly 
separating the $dU_L dU_R$ integration measure.
We can then compare $W(r,T) W(R,T)$ and $W \left( {R+r \over 2} , T 
\right)$.  A Schwartz inequality leads to
\eqa
W (r,T) W (RT) \geq  W^2 \left( {R+r \over 2} , T \right)
\eqae
which using (4) implies the convexity of the potential
\eqa
V \left( {R+r \over 2} \right) \geq {1 \over 2} [V (R) + V (r) ]
\eqae

Let us next prove the positivity of the Fourier transform  (F.T.) of 
$e^{-TV(r)}$.  We will do so by a slight
modification of the proof of the positivity of F.T. of certain 
four-point functions \cite{nine}.  Let us then consider
the following evolution which is a slight modification of that 
describe above in the proof of equation (4).  Let the
heavy quarks
$Q \bar{Q}$ be produced at $t=0$ and $\vec{r} =0$, and then separate, 
still at $t=0$, by moving $Q$ to $\vec{r}$ and
$\bar{Q}$ to $\vec{r}_2$.  After pure-timelike propagation to $t=T$ 
we bring back $Q$ and $\bar{Q}$ to $\vec{r}
=0$ and annihilate them there.  In this case the Wilson loop factor 
$W (\vec{r}_1 , \vec{r}_2, T)$ corresponds to the
closed path of fig. 1.

We can separate it into the product of $U$ matrices along the
``$\vec{r}_1$ like line" (going from $\bar{0}, 0$ to
$\vec{r}_1 , 0$, then to $\vec{r}_1 T$ and then to $0,T$) and a 
similar product over the reverse $\vec{r}_2$ line"
(connecting $\vec{0}, 0$ to $\vec{r}_2, 0$ to $\vec{r}_1, T$ to $\vec{0},T$)
\eqa
U = U (\vec{r}_1) U^\dagger (r_2)
\eqae
We are interested in the Fourier transform of $e^{-TV(R)}$.  To find 
the latter courier
\eqa
L(\vec{P}, T) \equiv \int\ d \vec{r}_1 d \vec{r}_2 e^{i \vec{p} ( 
\vec{r}_1 - \vec{r}_2)} <  W_c (\vec{r}_1,
\vec{r}_2 T) >\; .
\eqae
By exchanging orders of the trace, the $d \vec{r}_1, d \vec{r}_2$ 
integration and the $d \mu (U)$ path integral
averaging we can write:
\eqa
&& L( \vec{P}, T) = \int d  \mu (U) tr \Bigg[ \left( \int d\vec{r}_1 
U(\vec{r}_1) e^{i \vec{P} \vec{r}_1} \right)
\nn
&& \hspace{1.5in} \left( \int d \vec{r}_2 U( \vec{r}_2  ) e^{i 
\vec{P}_1 \vec{r}_2} \right)^\dagger \Bigg]
\eqae
which is manifestly positive definite.  Next we note that apart from 
some initial and final small corrections the
$Q$ and $\bar{Q}$ were at a fixed distance $\mid \vec{r}_1 - 
\vec{r}_2 \mid$ for all the  duration of the
evolution.  Hence we expect, particularly in the $T \rightarrow 
\infty$ limit of interest, that \cite{ten}

\eqa
W_c (\vec{r}_1, \vec{r}_2, T) \under{T \rightarrow \infty}{\approx} 
E^{-V (\vec{r}_1 - \vec{r}_2)T}
\eqae
The positivity of $L(P,T)$ then implies the desired positivity of the 
Fourier transform of $e^{-TV(r)}$.

\bigskip

The author would like to acknowledge the hospitality of the Department
of Physics of the University of Maryland.

\begin{figure}

  \begin{centering}
  \def\epsfsize#1#2{0.75#2}
  \hfil\epsfbox{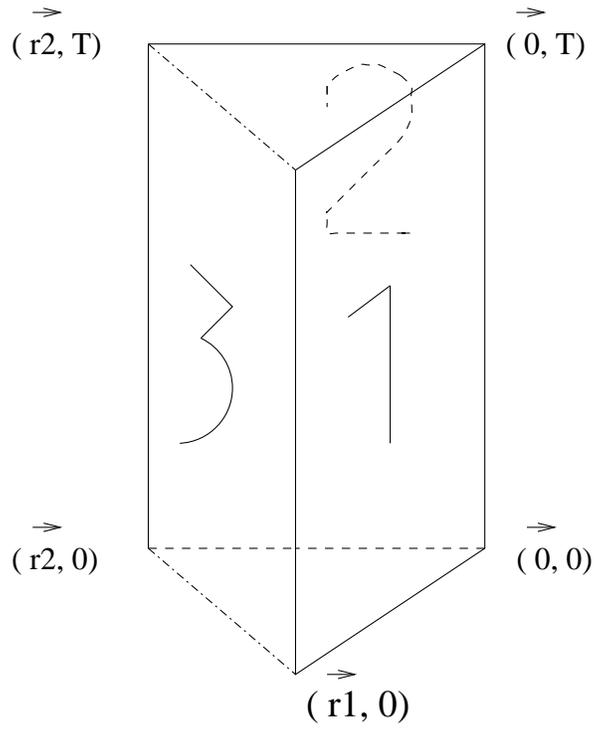}\hfil\hfill 

\bigskip

\caption{The location of initial $t=0$ and final $t=T$ heavy quarks
and the Wilson paths described by them.}

\label{tr}

\end{centering}
\end{figure}


\begin{references}
\bibitem{ten}  From confinement we expect that $<W_c> \propto 
e^{-A_c}$ with $A_c$ the area of the minimal
surface spanned by $c$.  In the case of fig. 1 this surface clearly 
does not consist of the rectangular faces 1 and 2
of area $\mid r_1 \mid T$ and $\mid r_2 \mid T$ respectively, but 
rather mainly of face the third face with area
$\mid \vec{r}_1 - \vec{r}_2
\mid T$, (recall that $\mid r_1 \mid + \mid r_2 \mid \geq \mid 
\vec{r}_1 - \vec{r}_2 \mid$ is the triangle
inequality).

The finite ``end-caps" effect translates here to contribution of the 
area of the two top/bottom triangles of sides
$\vec{r}_1, \vec{r}_2$ and $(\vec{r}_1 - \vec{r}_2)$ which are $T$ 
independent and irrelevant in the $T
\rightarrow \infty$ limit of interest.   We note however that these 
contributions proportional to $e^{-r_1{}^2 ( \;
{\rm or} \; r_2{}^2)}$ or to $e^{-M ( \vec{r}_1 - \vec{r}_2 \mid}$ 
ensure that the $d (\vec{r}_1 + \vec{r}_2)$
integration in (10) or (11), which does not pertain to the F.T. of 
$e^{-TV (\vec{r}_1 - \vec{r}_2)}$, converges.
\end{references}
\end{document}